# High throughput optimization of hard and tough TiN/Ni nanocomposite coatings by reactive magnetron sputter deposition

Ignacio Lopez-Cabanas[1,2], Javier LLorca[1,3], Raquel González-Arrabal[4,5], Efstathios I. Meletis[2] and Jon M. Molina-Aldareguia[1,*]

[1] IMDEA Materials Institute, c/Eric Kandel 2, 28906 Getafe, Madrid, Spain

[2] Department of Materials Science and Engineering, University of Texas at Arlington, TX, 76019, USA;

[3] Department of Materials Science, Polytechnic University of Madrid, E. T. S. de Ingenieros de Caminos, 28040 - Madrid, Spain

[4] Instituto de Fusión Nuclear "Guillermo Velarde", ETSI de Industriales, Universidad Politécnica de Madrid, Madrid, Spain

[5] Departamento de Ingeniería Energética, ETSI de Industriales, Universidad Politécnica de Madrid, Madrid, Spain

*Corresponding autor: jon.molina@imdea.org

**Abstract**

A combinatorial thin-film synthesis approach combined with a high throughput analysis methodology was successfully applied to synthetize TiN/Ni coatings with optimum mechanical properties. The synthesis approach consists of the deposition of TiN/Ni coatings with the Ni content ranging from 0 to 20 at.%. by reactive magnetron sputtering with a well-defined composition gradient, so that all possible compositions are produced in one single coating under identical conditions, removing uncertainties from processing conditions. The analysis methodology continuously screened the microstructure, residual stresses, hardness and fracture toughness of TiN/Ni coatings. The results show that a nanocomposite microstructure of equiaxed nanograins of crystalline δ-TiN embedded in a Ni rich amorphous phase is formed in the composition window between 8 and 12 Ni at.%. This microstructure offers a superior combination of hardness and toughness and a simultaneous reduction of residual stresses with respect to TiN coatings grown in the same conditions. Higher Ni contents are however detrimental because they compromise the integrity of the coatings.

## 1. Introduction

Since the pioneering work of Veprek and Reiprich [1], the concept of superhard nanocomposite nitride coatings has gained increasing attention. They are composed of hard nanocrystalline (nc) nitride grains embedded in an amorphous (a) matrix, forming a thin tissue of only a few monolayers thick around the nanocrystalline grains. Nanocomposite nc-TiN/a-$Si_3N_4$ coatings have been particularly studied and a hardness up to 50 GPa has been reported [2]. Such high hardness values are only obtained for narrow Si contents, between 5 and 10 at. %, and have been explained as a combination of grain refinement of the nc-TiN phase and the blocking of grain boundary sliding due to the strong interface that develops between the TiN grains and the $Si_3N_4$ amorphous tissue. Since then, these design principles have been extended to other coating systems, including TiAlN/$Si_3N_4$ [3], CrAlN/$Si_3N_4$ [4] and attempts have been even made to substitute Si by B [5]. In all of these cases, it was recognized that the thickness of the grain boundary tissue and its strength was critical to achieve a very high hardness. However, wear resistance does not only require a high hardness but a high toughness, i.e., a high resistance to cracking, should also be considered [6]. Systematic investigations of the fracture toughness of hard coatings are still scarce, but recent studies in nanocomposite CrAlN/$Si_3N_4$ coatings have shown an increased fracture toughness with respect to the monolithic CrAlN coating counterparts [7, 8]. This has been attributed to the formation of the nanocomposite microstructure triggered by the Si addition, which suppresses the brittle columnar boundaries responsible for the low fracture resistance of the latter. However, the role of the $Si_3N_4$ tissue on fracture resistance has not been investigated in detail. For instance, a complete blocking of grain boundary sliding might not be optimum to improve toughness because crack opening becomes the main mechanism to relieve strain if plastic dissipation is suppressed [6]. In this sense, one can anticipate that the use of a metallic tissue phase, more ductile than $Si_3N_4$ (such as, for instance, Ni, Co or Cu), could offer a high hardness, while improving toughness if the same nanocomposite structure is attained. Indeed, different attempts have been made to deposit composite TiN/Ni [9, 10], ZrN/Ni [11, 12], or ZrN/Cu coatings [13], but the range of compositions screened was limited. This is because, in most cases, the coatings were deposited from pre-alloyed targets with a fixed metal content, and finding the optimum composition would have been a tedious process requiring the deposition of a large number of coatings [9, 14, 15]. Moreover, the fracture toughness of such metal-ceramic nanocomposite coatings has not been systematically studied due to the lack of suitable techniques to measure the fracture toughness in thin-films. Instead, indirect methods, such as different hardness (H) to elastic modulus (E) ratios (H/E or $H^3/E^2$), or nanoscratch tests have been traditionally used to provide a qualitative assessment of the fracture resistance [16–18].

These limitations can be overcome by using a high-throughput technique based on combinatorial thin-film deposition [19] and by the exploitation of advanced characterization techniques with the required spatial resolution. Combinatorial synthesis of thin-films is one of the strategies that has been proposed to produce material libraries in multi-element systems in the last decade [20]. It involves the deposition of thin-films with a well-defined composition gradient and presents several advantages for material discovery and accelerated development of new coatings with respect to traditional approaches. Firstly, all possible compositions are produced in one single coating under identical conditions, removing uncertainties from processing conditions. Secondly, the composition space can be continuously screened owing to the development of novel nanomechanical characterization techniques that allow not only to interrogate the local hardness and modulus by fast mappings, but also to locally measure other important coating properties, like residual stresses [21] or fracture toughness [22]. In this way, optimum combinations cannot be missed.

In this work, this high-throughput characterization strategy has been applied to TiN/Ni coatings where the Ni content continuously varied between 0 and 20 at.%. The elemental composition and structure of the coatings was spatially mapped by combining Rutherford Backscattering Spectroscopy (RBS), X-Ray energy dispersive spectroscopy (EDS) and X-Ray Diffraction (XRD). The hardness and modulus were determined by high-speed mapping using instrumented nanoindentation, while the residual stresses and fracture toughness were locally measured in selected locations by focused ion beam (FIB) milling combined with digital image correlation (DIC) and micropillar splitting tests, respectively. Thus, the effect of metal content on the mechanical properties of TiN/Ni composite coatings has been systematically studied and correlated to the coating microstructure by using transmission electron microscopy (TEM). A nanocomposite microstructure with optimum mechanical properties was found when the Ni content was in the range 8 to 12. at %.

## 2. Experimental procedure

A reactive magnetron sputtering system, built in-house, was used for the synthesis of the coatings. The coatings were deposited on a long strip of a Si <100> single-crystal, with dimensions 10 × 100 mm, by co-deposition from Ti (99.95%) and Ni (99.9%) targets. The base pressure of the sputtering chamber was $3.4 \cdot 10^{-4}$ Pa. The atmosphere in the chamber was composed of Ar with 15% $N_2$ and the pressure during deposition was set to 0.66 Pa.

Suitable samples for testing the combinatorial synthesis approach must present a compositional gradient along them. To achieve this, the holder was maintained stationary during deposition, so that the composition of the coating along the long edge of the substrate varied as a function of the distance to each target. As shown in Figure *1*, the Si strip was placed in such a way that one edge of the substrate was directly facing the Ti target, while the rest of the strip was aligned with the long edge following the diameter of the substrate holder. The distance between the targets and the substrate holder was 4" (~10cm), which was the holder's diameter as well. The targets' diameter was 2" (~5cm). Prior to the deposition of the coatings, selected areas of the substrate were masked in order to determine the coating's thickness using optical profilometry. For the sake of easy identification of the location of the different measurements along the compositional gradient, the coating was structured in different measurement areas, of size around 10 x 20 mm, designated as TiN/Ni-1 to TiN/Ni-5, where 1 and 5 represent the locations that were the closest and the furthest, respectively, to the Ti target.

The targets were initially sputter cleaned in Ar. The substrate temperature was set to 400º C in vacuum. Not rotating the holder might lead to thickness uniformity and temperature homogeneity issues that should be taken into account. A 45 min thermal stabilization stage was initially carried out, followed by the deposition of a 50 nm thick Ti interlayer, to avoid large thermal gradients and reduce the associated thermal stresses. The interlayer deposition was carried in an Ar pressure of 0.66 Pa with the Ti target operated at 150 W and using holder rotation at 30 rpm to ensure the uniformity of the interlayer. Then, substrate rotation was stopped, $N_2$ flow was allowed aiming at a $N_2$ partial pressure of 15% and maintaining the total pressure at 0.66 Pa, and the Ni target was switched on. The Ti and Ni targets were operated at a power of 150 W and 12 W, respectively, and the substrate bias was held at -100V during deposition. The power applied was calibrated so that the deposition rates met a 10:1 ratio at the center of the holder. The deposition was extended aiming at a total thickness of approximately 1 μm. However, the final total thickness of the coatings was 750 nm on average. The thickness was uniform along the Si strip within 10%, as measured by optical profilometry from 7 different positions spaced approximately 1 cm. After deposition, the coating was cooled down in a nitrogen atmosphere in order to avoid its oxidation.

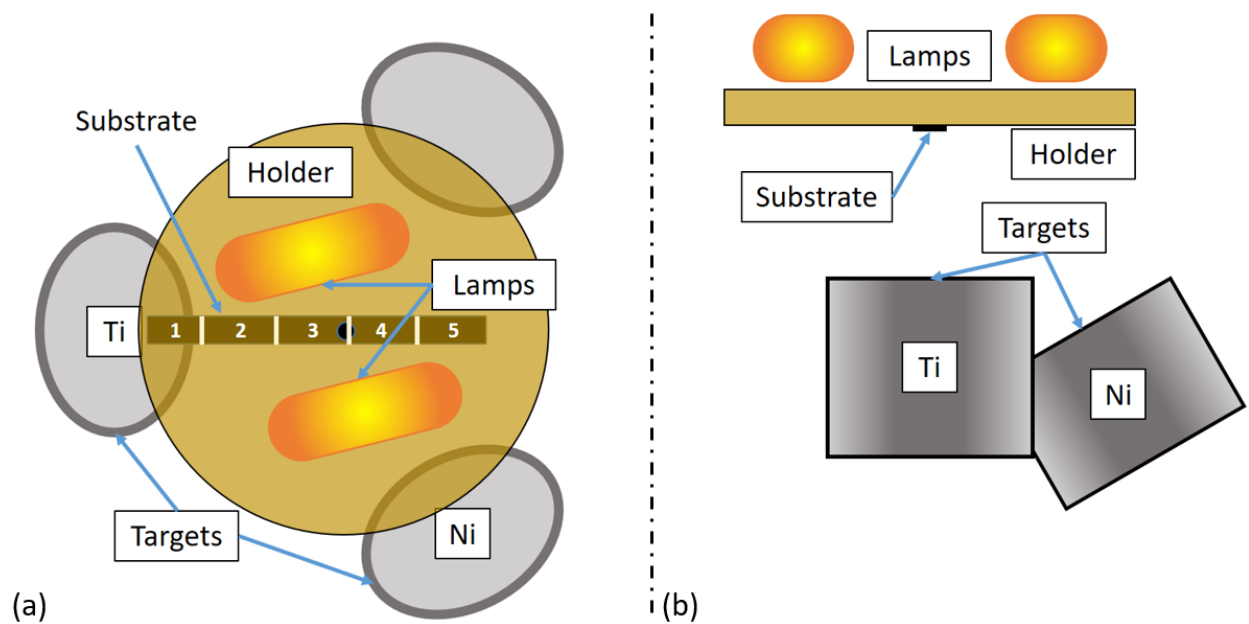

Figure 1. (a) Top view and (b) side view schematic illustrating the deposition set up and the relative position of the substrate, and the corresponding measurement areas (1 to 5), with respect to the Ti and Ni targets.

A range of locally resolved characterization techniques were used to map the elemental composition, structure and mechanical properties of the TiN-Ni coatings as a function of location along the substrate. These techniques have different lateral resolutions. Additionally, some can be more easily applied following a high-throughput strategy than others. Therefore, some characterization techniques were applied continuously with a high lateral resolution along the substrate position, while others were only used in selected measurement areas. These considerations must be taken into account for interpretation and comparison of results from different characterization techniques.

The elemental composition was spatially mapped by RBS and EDS. High-throughput EDS was used to estimate the Ni content and the Ni/Ti ratio along the position of the coating with a high resolution (interaction volumes are of the order of a few micrometers) and a high precision (typically ±1%). Measurements were taken with a step size of 1 mm along the long edge of the substrate. However, EDS is not accurate enough for light elements, so non-RBS measurements were carried out in selected locations, to determine the N content, derive the stoichiometry of the TiN phase and validate the high-throughput EDS measurements. The RBS measurements were performed in a high vacuum chamber connected to a 2 MV Tandetron accelerator [23], using a He beam at an energy of 3.7 MeV and a spot size of the order of 1.5 mm. At this energy the cross-section for N become non-Rutherford which improves the resolution accuracy of the technique for the determination of the elemental concentration of N [24]. The backscattered ions were detected by a standard Si-barrier detector. The composition was estimated by comparing experimental and simulated spectra following the procedure described in [25]. The commercial computer code SIMNRA version 6.02 was used for the simulations [26].

Phase analysis of the samples was done by XRD using a PANanalytical P4311 diffractometer in θ/2θ configuration using Cu Kα radiation. The measurement area was around 5×5 mm, and therefore, the XRD spectra covered areas where the elemental composition could not be considered uniform, but gradually changed. Despite the poor lateral resolution, this was considered a good compromise to qualitative assess the evolution of the structure of the coatings with composition, because smaller irradiated areas resulted in too low intensities in the XRD spectra.

Residual stresses, inherent to any coating process, are also important, because they can contribute to improve the hardness, but also compromise the coating adhesion. Residual stresses arise as a result of the complex interplay between deposition conditions, specially ion bombardment effects and temperature of deposition, microstructure evolution and the constraint imposed by the substrate. Traditional approaches to measure residual stresses, either by XRD diffraction or substrate curvature measurements, lack the required spatial resolution to be applied to coatings with a spatial compositional gradient. In this work, the methodology proposed by Korsunski *et al.* [21] to determine residual stresses at the microscale was used as it offers the required lateral resolution to map residual stresses in different locations of the coating along the compositional gradient. The methodology involves the incremental FIB milling of annular trenches, combined with tracking the position of previously formed markers using high resolution scanning electron microscopy (SEM) images, to assess the strain relief upon material removal [21]. To this end, a Pt layer was FIB deposited over selected areas of the coating and a pattern of holes was milled by FIB to serve as the marker pattern, as shown in Figure *2* (a) [21]. Afterwards, the diameter of the annular trenches was selected to be equal to the film thickness and they were milled to the same depth, to ensure total strain relief of the coating and to allow the determination of the average residual stress in each location, as shown in Figure *2* (b). Due to the small dimension of the annular trenches, it was critical to ensure that the SEM images were taken at the maximum resolution and were free from artifacts, such as drift, that can compromise the tracking of the surface pattern. Thus, the coatings were gold sputtered to improve the surface conductivity and minimize drift during image acquisition. The SEM images were taken with a resolution of 6114 × 4415 pixels and a dwell time of 1 μs using three different scanning rotations of 0, 45 and 90º degrees with respect to the marker pattern. In this way, strain relief could be determined along three different directions in each case but ensuring that the measurements between the markers were always taken in the scanning direction of the electron beam, which is less susceptible to drift artifacts. The residual stress ($\sigma_R$) can be finally determined from the average strain relief ($\bar{\epsilon}$) assuming a state of biaxial stress according to:

$$\sigma_R = \frac{E}{1-\nu}\bar{\varepsilon} \qquad (1)$$

where $E$ is the elastic modulus and $\nu$ is the Poisson ratio of the coatings.

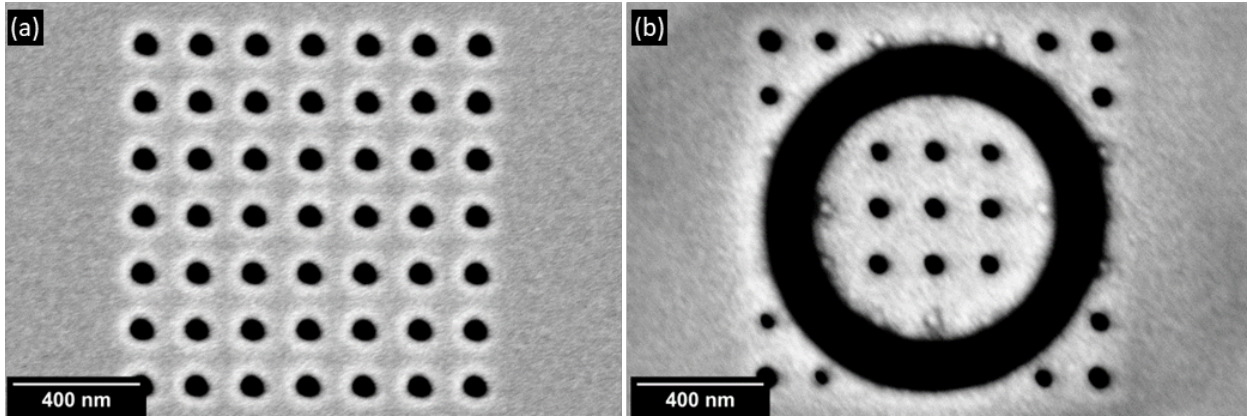

Figure 2. Residual stress measurement steps, (a) pattern mill and (b) micropillar mill.

The hardness and elastic modulus were determined by high-speed instrumented nanoindentation mapping with a Hysitron Triboindenter TI950 system equipped with a Berkovich diamond tip. Nanoindentation tests were performed in depth control at a maximum depth of 70 nm to avoid substrate effects. Indentations were made in approximately 150 locations, spaced 325 µm along the long edge of the specimen, to cover the entire composition profile using a high-throughput approach. The results were then correlated to the composition profile determined by EDS. Hardness and elastic modulus were calculated using the Oliver and Pharr method [27] assuming a Poisson ratio for the coating of $\nu$=0.24 [28]. Four indents were made at each location, spaced 325 µm along the direction perpendicular to the compositional gradient, to determine the average properties and the standard deviation as a function of chemical composition.

As discussed in the introduction, the fracture resistance of the coating should also play a major role on the wear resistance of the coatings. This factor is particularly important in this work where one of the main objectives is to elucidate whether the incorporation of a metallic phase into nitride hard coatings can potentially contribute to fracture toughness enhancements without sacrificing the hardness. The micropillar splitting technique proposed by Sebastiani *et al.* was used [7]. This technique can hardly be considered a high-throughput approach as it requires the FIB milling of micropillars, which is time consuming. However, it allows enough spatial resolution, of the order of the micropillar diameter, to map the fracture toughness of the coating in selected locations along the compositional gradient of the coating. Grids of 4 micropillars were prepared in 12 different locations along the direction of the compositional gradient, with a spacing between locations of around 5 mm. The spacing between micropillars within each grid was 10 µm, as shown in Figure *3* (a). Therefore, each grid represented a fixed chemical composition, so that the average and standard deviation of the measurements from each location could be computed as a function of chemical composition. The micropillars were FIB machined using a FIB-FEGSEM dual-beam microscope (Helios Nanolab 600i) to a diameter equal to the film thickness, of the order of 1 μm. as shown in Figure 3 (b). They were indented using the same nanoindentation instrument, but equipped with a cubic corner diamond tip. The fracture toughness $K_{IC}$ was determined from the critical load for micropillar splitting, $P_c$, using:

$$K_{IC} = \gamma \frac{P_c}{R^{3/2}} \qquad (2)$$

where $R$ is the radius of the micropillar and $\gamma$ is a factor that depends on the indenter geometry and the *H/E* ratio of the coating and that was obtained from the work of Ghidelli *et al*. [29]. In this case, $\gamma$ took values in the range 0.63-0.83.

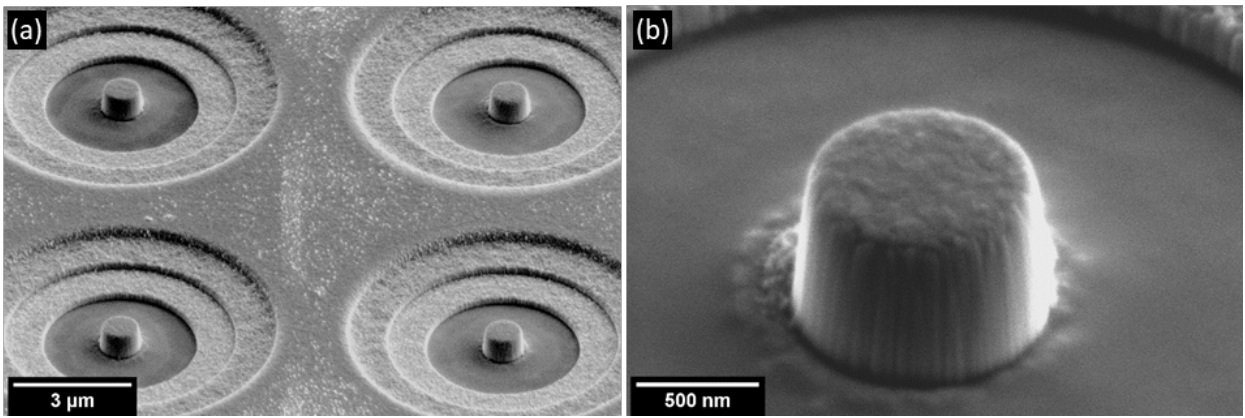

Figure 3. (a) Grid of micropillar milled for pillar splitting experiments; (b) High magnification image of one of the micropillars.

Finally, the microstructure was analyzed by scanning transmission electron microscopy (STEM) using a Thermosfisher Talos F200X microscope equipped with EDS to get a better understanding of the changes observed in the mechanical properties with the elemental composition of the TiN/Ni coatings. The TEM specimens were prepared in selected locations by the lift-out technique using the same FIB-FEGSEM dual-beam microscope (Helios Nanolab 600i).

## 3. Results

*3.1 Elemental composition*

The Ni content is plotted in Figure *4*, measured by EDS as a function of position along the Si strip. The origin represents the location in front of the Ti target. The EDS measurements show that a coating with Ni contents ranging from <0.5 at. % to approximately 20 at. % along a distance of 90 mm was successfully deposited. This is a perfect interval to evaluate the impact of Ni addition on the structure and mechanical properties of TiN coatings since the origin stands for an almost pure TiN coating that can be used as reference, even though small amounts of alloying elements can have some impact on the mechanical properties due to grain refinement effects [30]. The dash vertical lines in Figure *4* represent the measurement areas in which the coating was structured to aid for the easy identification of the location of the different measurements along the compositional gradient. The bulk of the variation in Ni content occurred in locations close to the center of the substrate holder, along the measurement areas TiN/Ni-2 to TiN/Ni-4, while the variations on the edges of the long strip were smaller. In fact, the TiN/Ni-1 and TiN/Ni-5 measurement areas were not mapped with a high spatial resolution, because the Ni content did not vary as much as in the rest of the sections.

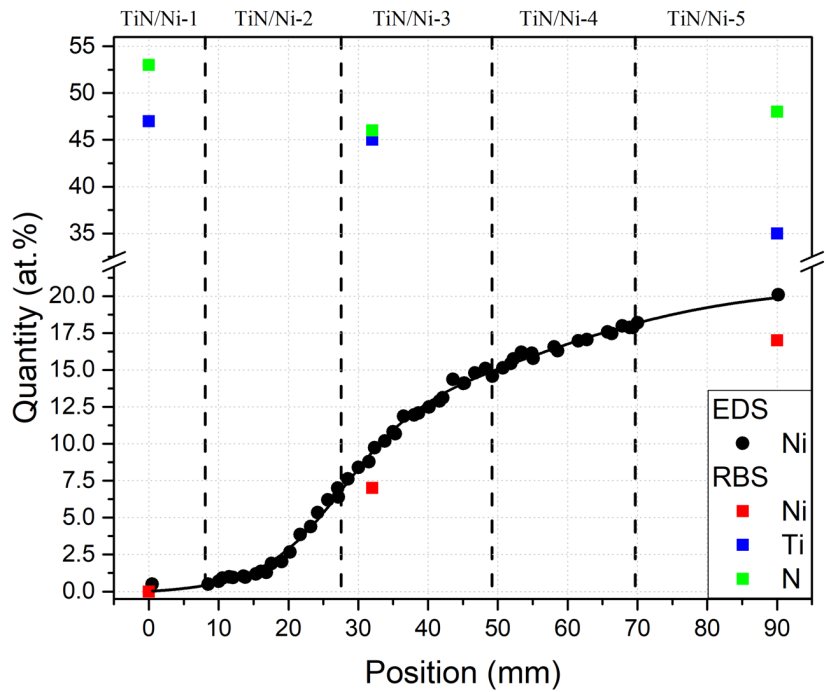

Figure 4. Ni, Ti and N (RBS) and Ni (EDS) content along the sample.

Since EDS is not accurate enough for the quantification of light elements like N, RBS measurements were carried out in three selected locations of areas TiN/Ni-1, TiN/Ni-3 and TiN/Ni-5, to determine the N content and to validate the EDS results. The results are also included in Figure 4. The Ni content determined by EDS and RBS agreed reasonably well. EDS overestimated the Ni content by approximately 10%, which is not surprising considering that the EDS measurements are influenced by the uncertainty in the N content. More interestingly, the RBS results showed that the N content remained constant at around 50 ± 4 at. %, along the entire Si strip. In other words, the (Ti+Ni)/N ratio was around 1 ± 0.1 along the whole coating.

More specifically, the Ti/N ratio was 0.9 in the pure TiN side indicating an almost stoichiometric TiN compound, with just a small excess of N, corresponding to TiN$_{1.1}$. This is important because TiN$_x$ can be stable over a wide range of stoichiometries, with *x* ranging from 0.6 to 1.2 [31], and reactive sputtering can yield TiN$_x$ films with a wide range of Ti/N ratios depending on the deposition conditions. The RBS results confirm that stoichiometric TiN was achieved under the sputtering conditions used in this work, at least in the part of the coating that was located in front of the Ti target. Nevertheless, the fact that the (Ti+Ni)/N ratio remains constant at around 1 ± 0.1 as Ni incorporates into the film might indicate that Ni dissolves into the TiN phase or that the Ni rich phase might contain a substantial amount of N if segregated out as a Ni rich phase. In fact, TiN and Ni are immiscible and Ni is not a strong nitride former [32], but sputtering conditions are far from equilibrium and co-deposition of Ti and Ni in the presence of nitrogen can potentially result in any atomic mixture in the as-deposited film.

*3.2 Structural characterization*

The XRD spectra from the different TiN/Ni measurement areas in the as-deposited state are depicted in Figure *5*. The Ni composition varies within each measurement area in the ranges indicated on the right side of the figure. This might affect the XRD analysis, as the microstructure is not expected to be homogeneous within each measurement area. The only crystalline phase present in the coatings was δ-TiN, independently of Ni content, indicating that either Ni dissolves within the crystalline δ-TiN phase or segregates to form a Ni rich amorphous phase during deposition. Starting with the region TiN/Ni-1, with contents below 0.5 at. %Ni, the δ-TiN phase grows with a (111) preferred growth orientation and the TiN (200) peak is insignificant. A transition from a (111) to a (200) preferred growth orientation was observed in regions TiN/Ni-2 and TiN/Ni-3, as the Ni content increased up to 15 at.%. However, the δ-TiN did not show any preferred crystallographic texture for contents larger than 15 Ni at. %. Finally, , the δ-TiN peaks shifted to higher diffraction angles with respect to the reference peak positions (indicated by the vertical lines) and their full width at half maximum (FWHM) increased with increasing the Ni content. The latter is indicative of a smaller crystalline domain sizes or of a reduction in the crystallinity of the coatings as the Ni content increases. The former indicate changes in the out-of-plane lattice parameter of the δ-TiN phase, which might arise as a result of residual stresses and/or changes in its chemical composition and will be discussed in more detail below.

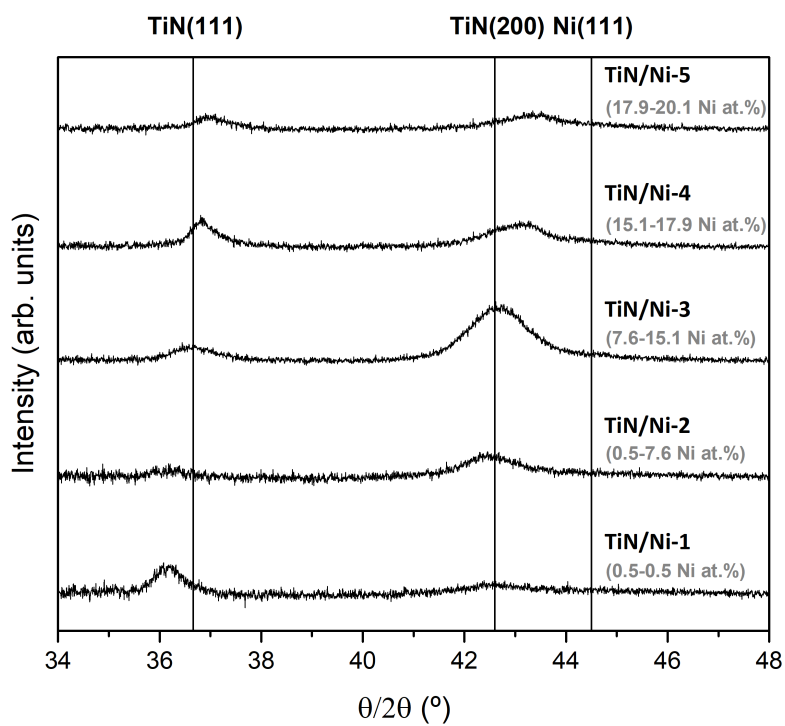

Figure 5. XRD patterns of the TiN/Ni coating along the different measurement areas.

*3.3 Mechanical properties and residual stresses*

The hardness, elastic modulus and fracture toughness mapping are plotted in Figure 6 as a function of Ni content. The elastic modulus was fairly constant, around 340 GPa, up to 13 Ni at. % and dropped quickly for higher Ni contents. Both hardness and fracture toughness increased slightly with Ni content, from 25 GPa and 4.3 MPa·m$^{1/2}$, respectively, for very small Ni contents, up to 32 GPa and 4.8 MPa·m$^{1/2}$, respectively, for contents of the order of 10 Ni at. %. Both hardness and fracture toughness decreased slowly with a further increase in Ni content and dropped rapidly, together with the elastic modulus, for Ni contents > 13 at.%.

The in-plane biaxial residual strain (measured by digital image correlation upon FIB material removal) is also plotted on Figure 6 in different locations of the coatings. The measurements were carried out in three different in-plane orientations, as explained in the experimental section. The results confirmed a state of biaxial residual stress, as expected for thin films, as the strain relief estimated in the three different orientations provided similar values, within experimental error. The almost pure TiN coating was subjected to large compressive residual strains, of the order of 1.5%. Incorporation of 4 at. %Ni led to a dramatic reduction of compressive residual strains, down to 0.25%, while the residual strains were found negligible or even slightly tensile for Ni contents higher than 10 at.%.

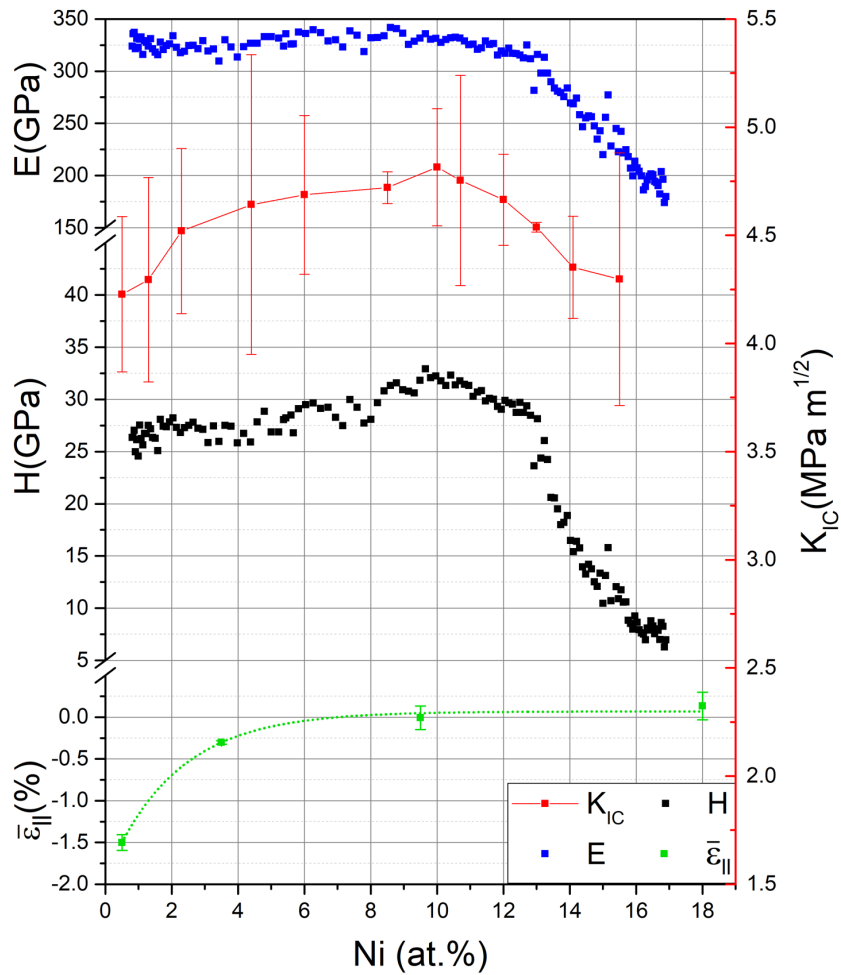

Figure 6. Hardness, elastic modulus, in plane biaxial residual strain and toughness as a function of the Ni content.

*3.4 Microstructure*

The microstructure of the TiN-Ni coatings as a function of Ni content was analyzed using TEM . Electron-transparent lamellae were extracted by FIB from three different locations corresponding to pure TiN, TiN-10 at.% Ni and TiN-18 at.% Ni. A bright-field (BF) image of the cross-section of the TiN coating is shown in Figure 7 (a). The Ti interlayer can be clearly seen, as well as the development of a columnar grain microstructure. The columnar grains have an inverted pyramidal shape which is compatible with the zone T in the Thornton's coating morphology diagram [33]. The columnar widths increase with coating thickness and reach tens of nanometers at the coating surface. The diffraction patterns (DP) in Figure 7 (b) and (c) show a transition from more diffuse δ-TiN (111) and (200) diffraction rings to distinct diffraction spots with coating thickness, as a consequence of the increasing width of the columnar grains. Figure 7(d) shows a high-resolution electron microscopy (HREM) image, demonstrating a dense microstructure, without the presence of voids or pores along the columnar boundaries. This is the typical microstructure found in sputtered dense stoichiometric TiN coatings that develops as a result of competitive growth between different grain orientations in kinematically limited conditions [34], which leads to the (111) preferred orientation determined by XRD (Figure 5).

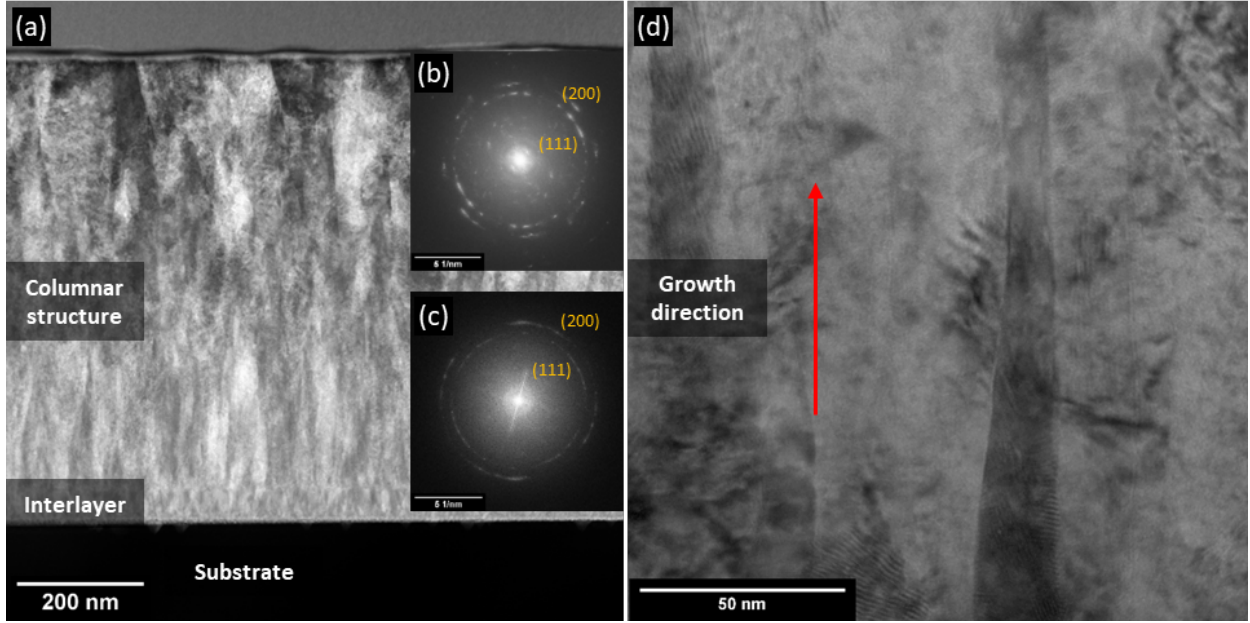

Figure 7. Cross-sectional images of the TiN coating: (a) BF TEM; (b) and (c) DPs taken in the marked locations; (d) HREM showing the corresponding lattice fringes and dense columnar boundaries

The TEM images corresponding to TiN/Ni with a nickel content of 10 at.% (with maximum for which hardness and fracture toughness) are depicted in Figure 8. The BF image in Figure 8(a) shows that the columnar structure gradually transforms into a nanocrystalline structure with equiaxed grains as the coating grows. This transition is likely due to the time required during sputter deposition to reach the prescribed 15% partial pressure of $N_2$ after opening the $N_2$ flow. The nanocrystalline nature is clearly demonstrated by the diffuse diffraction rings in the DP taken in this region (Figure 8(b)) and by the HREM image i Figure 8(d), which shows equiaxed crystalline nanograins, with sizes of the order of 5-10 nm. Finally, the presence of an amorphous phase in some regions is also evident in Figure 8(d), as pointed by the arrow. High-angle annular dark field (HAADF) images (Figure 8(e)) and EDS maps (Figure 8(f)) in the same region also demonstrate that Ni is strongly segregated in this area, forming small domains that are evenly distributed at the same scale than the nanograin size without any evident directionality. TEM is a projection technique and the images show the superposition of several nanograins along the thickness of the TEM foil, which is of the order of 100 nm. Therefore, it is impossible to determine the chemical composition and spatial distribution of the crystalline and amorphous regions, but the TEM observations, combined with the XRD results (Figure *5*), are compatible with what is expected for a nanocomposite microstructure formed by crystalline nanograins of δ-TiN, preferably oriented in the (200) growth orientation, embedded in a Ni rich amorphous phase.

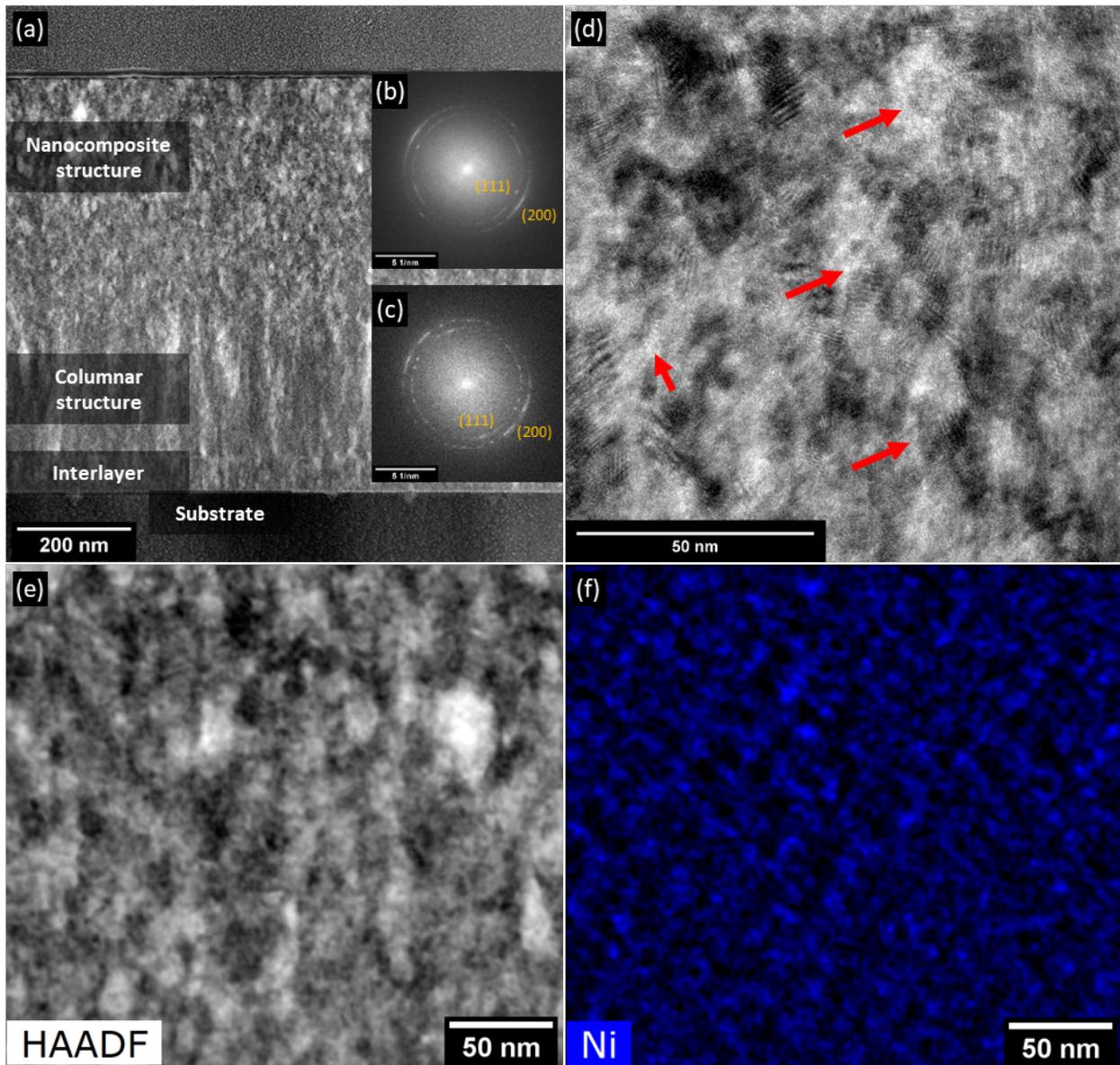

Figure 8. Cross-sectional images of the coating on the location corresponding to TiN with a Ni content of 10 at. %: (a) BF TEM image; (b) and (c) DPs taken in the marked locations; (d) HREM, showing small equiaxed crystalline grains surrounded by an amorphous phase; (e) HAADF STEM and (f) corresponding EDS map showing the Ni distribution

TEM images of TiN/Ni with 18 at.% Ni are shown in Figure 9. The nanocomposite structure is lost with this Ni content, as shown by the distinct diffraction spots found in the DPs of Figure 9 (b) and (c). In fact, the BF image in Figure 9 (a) shows that the microstructure is composed of columnar grains but, contrary to the TiN rich location, the columnar grains are narrower, the coating develops a relatively rougher growth front and the columnar grain boundaries are more defective.

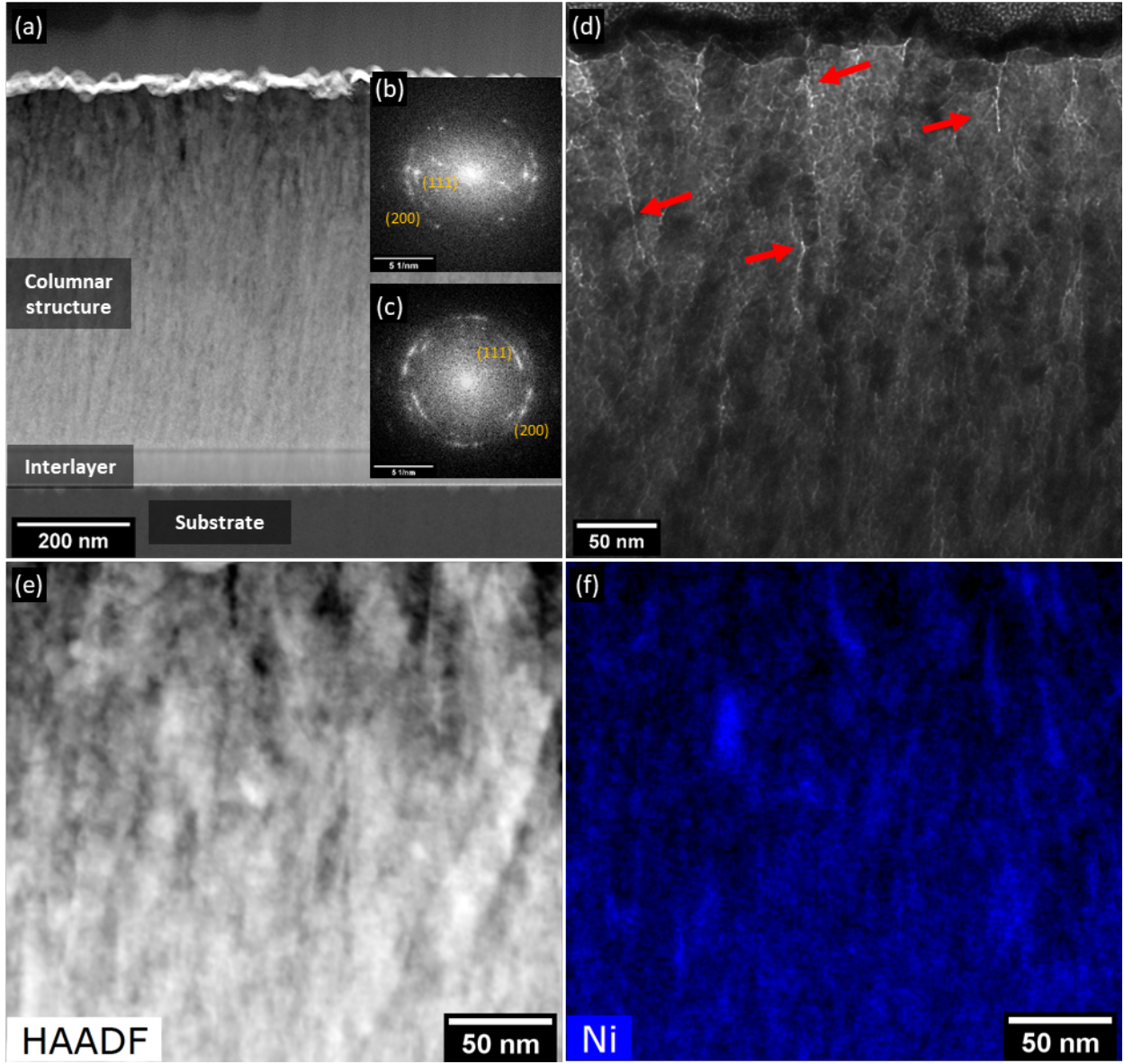

Figure 9. Cross-sectional images of the coating on the location corresponding to TiN with a Ni content of 18 at.%: (a) BF TEM image; (b) and (c) DPs taken in the marked locations; (d) BF TEM under slightly defocus conditions, showing defective columnar boundaries; (e) HAADF STEM and (f) corresponding EDS map showing the Ni distribution

In fact, the presence of a relative large fraction of voids along the columnar boundaries is evident from the bright lines that appear in the BF image of Figure 9 (d), taken under slightly underfocused conditions to enhance Fresnel contrast. Finally, the EDS Ni maps (Figure 9 (f)), corresponding to the HAADF STEM image in (e), show that Ni segregates into relatively large veins preferably oriented along the growth direction.

## 4. Discussion

The results of this investigation demonstrate that Ni addition during growth of TiN coatings by reactive sputtering can trigger the formation of a nanocomposite structure formed of equiaxed nanograins of crystalline δ-TiN embedded in a Ni rich amorphous phase. This microstructure shows a superior combination of hardness and toughness without residual stresses than can be beneficial for wear applications. Similar results have been observed for TiN/Ni coatings in previous works [9, 14, 35], but only for a limited number of chemical compositions, and using special deposition setups that promote strong ion irradiation conditions, like cathodic arc evaporation [36] or ion beam assisted sputter deposition [10, 35]. As a result, the prevailing ion irradiation effects in such conditions could not be completely decoupled from elemental composition effects on the microstructure development. Moreover, most previous works only measured the hardness of TiN/Ni composite coatings, but the impact of the microstructure on the fracture toughness of the coatings was unknown. The combinational thin-film synthesis approach used in this work, together with the use of advanced nanomechanical testing techniques, allowed the continuous screening of the effect of Ni content on the microstructure, residual stresses, hardness and fracture toughness of TiN/Ni coatings. Moreover, all possible compositions are produced in one single coating under identical deposition parameters (sputtering gas, target powers, substrate bias, substrate temperature, etc.) removing urcerainties due to differences in processing conditions, and allows focus the discussion on the role of Ni addition.

Sputtered δ-TiN coatings are one of the most thoroughly studied hard coatings in the last decades and their stoichiometry, microstructure and mechanical properties are very dependent on the deposition conditions. According to Figure *4*, the δ-TiN phase in the TiN rich side was slightly over stoichiometric under the deposition conditions used in this work. Moreover, high residual compressive stresses appeared (Figure 6) and a (111) preferred growth orientation developed (Figure 5). The high compressive stresses generate due to atomic peening and the generation of defects caused by ion bombardment effects, inherent to the sputtering process. Texture development in TiN coatings is affected by many process parameters. It is generally accepted that the (111) preferred orientation arises through a competitive columnar growth process under kinematically limited conditions, because the (111) oriented columnar grains, which are the grains with the fastest growing direction, overlap the rest of the orientations [34]. The microstructure observed in this coating (Figure 7) agrees well with that expected through a competitive columnar growth. Moreover, the hardness and elastic modulus, 25 GPa and 330 GPa, respectively, are those expected for good quality (111) oriented dense TiN coatings [37].

As Ni was incorporated into the film, a clear transition was observed from a TiN (111) to a TiN(200) preferred growth orientation (Figure 5) for Ni contents up to 10-15 at. %, together with a complete relief of the compressive residual stresses (Figure 6). In this range of Ni contents, the microstructure transformed from a δ-TiN columnar grain structure (Figure 7) to an equiaxed δ-TiN nanocrystalline microstructure (Figure 8). The TEM studies suggest the segregation of an amorphous Ni rich phase in this case. However, it was not clear if under the non-equilibrium conditions found during sputtering, some Ni could also be dissolved within the δ-TiN phase. In this respect, it is interesting to note that the elastic modulus of the TiN/Ni composite coating remained constant at 330 GPa, up to 13 at.% Ni, while the hardness and fracture toughness peaked at 10 at. %Ni and decreased slowly up to 13 at. %Ni. Only did both the hardness and elastic modulus drop dramatically for larger Ni contents.

The elastic modulus is a property that depends on the strength of the atomic bonds in the coating, and it is relatively insensitive to minor additions of alloying elements and/or the material microstructure, as long as the material remains dense and the bond strength remains unaffected. This is particularly relevant for δ-$TiN_x$, which exists over a wide range of Ti/N ratios (0.6<x<1.2) because changes in composition can be accommodated by a high vacancy concentration in the nitrogen sub-lattice (for x<1) and the Ti sublattice (for x>1). However, such high levels of vacancy concentrations lead to strong reductions in elastic modulus and hardness in TiN coatings due to the impact of stoichiometry on bond strength [37]. The fact that the elastic modulus remained constant at 330 GPa up to 13 at.% Ni indicates, therefore, that the δ-TiN remained close to stoichiometry up to that point and enforces the conclusion that most of the Ni segregated out during the growth of the δ-TiN crystalline grains. As a matter of fact, the RBS results in Figure 4 indicate that the Ti/N ratio remained close to 1 at least up to 10 at.% Ni. Another indirect evidence of the exact chemical composition of the δ-TiN phase might be found on the value of the lattice parameter that can be extracted from the diffracted peaks in the XRD spectra that show substantial shifts with respect to the reference peak positions with Ni content (Figure 5). The lattice parameters calculated using Bragg's law in the Bragg-Brentano geometry provide the out-of-plane lattice parameter, $a_\perp$, of the diffracting grains, which might be affected, by chemical composition and by the in-plane residual stresses due to Poisson effects. It is, however, possible to estimate the stress-free lattice parameter, $a_0$, from the knowledge of the in-plane residual strain $\varepsilon_{//}$ measured in Figure 6 according to:

$$a_0 = \frac{a_\perp}{(1-2\upsilon\varepsilon_{//})} \qquad (3)$$

where $\nu$ is the Poisson's ratio of TiN. The results are plotted in Figure 10, which shows the measured out of plane lattice parameter as well as the estimated stress free lattice parameter from the (200) and (111) peaks as a function of Ni content. The horizontal dash line represents the reference lattice parameter for stoichiometric stress-free δ-TiN. The initial section with Ni contents below 0.5 at. % (TiN-1) reveals a slightly increased stress-free lattice parameter. Even though the maximum lattice parameter is expected for the stoichiometric compound [38], incorporation of N atoms in tetrahedral interstitial sites has also been observed for over-stoichiometric films (x>1) in the case of TiN films grown under non-equilibrium conditions, leading to an expansion in the lattice parameter [37], in agreement with our observations. Moreover, the high density of defects resulting from ion bombardment induced atomic peening might also contribute to this lattice parameter expansion [34]. With increasing the Ni content and with the transition to a (200) preferred orientation, however, the lattice parameter of the δ-TiN phase continuously decreased and reached the reference value for stoichiometric δ-TiN for Ni contents up to around 15%. The Ti/N ratio remained close to 1 up to at least 10 at.% Ni according to the RBS results, and this result supports the conclusion that most of the Ni segregates out of the TiN grains during growth. It appears, therefore, that the transition from a competitive columnar grain growth to a nanocrystalline nanostructure is triggered by the amorphous Ni phase stopping the growth of the TiN grains and forcing their re-nucleation, as it was proposed by Abadias [34], but without affecting their stoichiometry. This explains the development of a (200) preferred growth orientation, which is the preferred nucleation orientation of TiN because its surface energy is the lowest for the (001) surface [39], and the little variation of elastic modulus in this composition range. In the light of the hardness and fracture toughness results (Figure 6) the TiN/Ni interfaces that develop are much stronger that the columnar grain boundaries that develop in pure TiN coatings, which explains the continuous increase in hardness and toughness, even when the residual stresses are dramatically reduced.

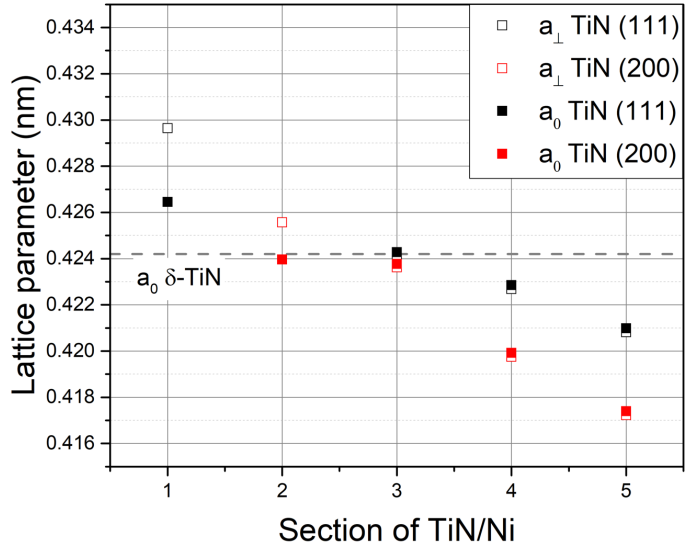

Figure 10. Evolution of the out-of-plane lattice parameter and estimation of the stress-free lattice parameter of the δ-TiN phase with Ni content

The dramatic reduction in compressive residual stresses with the incorporation of Ni also deserves further attention. One of the benefits of using the combinatorial approach presented in this work is that the compositionally graded film is grown under identical deposition conditions (bias voltage, magnetron power, gas flow rate, nature of sputtering gas species, temperature and pressure), and therefore, equivalent energy and flux of the bombarding species. If the compressive residual stresses found in the TiN rich side are due to atomic peening induced by ion bombardment, it is surprising that the incorporation of Ni to just a few atomic per cent produces such a drastic reduction of residual stresses. It is well known that the development of a (002) texture can reduce the defect density induced by ion bombardment due to channeling effects, because<001> is the most open channeling direction in TiN [40]. But, on top of texture effects, such a drastic reduction in compressive residual stresses can only be explained if the Ni rich amorphous phase that surrounds the TiN nanograins contributes to relieve part of these residual stresses.

Finally, the elastic modulus, hardness and fracture toughness of the coatings were drastically reduced for Ni contents higher than 13 at.% Ni. This reduction is triggered by the loss of the nanocomposite type microstructure. Instead, the microstructure evolves to the formation of columnar grains with a random texture surrounded by Ni-rich amorphous islands. Moreover, voids appear at grain boundaries (Figure 9), contributing to the large drop in hardness and fracture toughness. However, the reduction in elastic modulus can only be understood if the bonding strength of the TiN phase gets also compromised. As a matter of fact the lattice parameter of the δ-TiN phase for this Ni content shows values that are well below what is expected for stoichiometric TiN (Figure 10). It is hypothesized that either the TiN becomes highly over-stoichiometric or some Ni atoms get incorporated into the TiN phase by substitution of Ti atoms by Ni. This would result in a reduction in the lattice parameter because the ionic radius of $Ni^{3+}$ (0.056 nm) [41] is smaller than that of $Ti^{3+}$ (0.075 nm) [42], as has also been observed in $TiN/Ni/Si_3N_4$ composite coatings [43]. This hypothesis is supported by the fact that the N/Ti ratio is of the order of 1.4 for a Ni content of 17 at.% (Figure 4), while the (Ti+Ni)/N ratio remains constant at around 1. Considering that some of the Ni is present in the form of Ni rich amorphous islands, such atomic ratios can only be understood if the δ-TiN phase contains substantial amounts of Ni and/or a high fraction of vacancies in the Ti sub-lattice. In either case, this would result in a large reduction in the bond strength of the TiN phase.

## 5. Conclusions

A combinatorial thin-film synthesis approach combined with a high throughput analysis methodology was successfully applied to continuously screen the microstructure, residual stresses and mechanical properties of TiN/Ni coatings produced by reactive magnetron sputtering with Ni contents ranging from 0 to 20 at.%. The following conclusions were obtained:

- The optimum composition window to produce TiN/Ni coatings with maximum hardness and fracture toughness using standard reactive magnetron sputtering lays in the range 8 to 12 at. %.
- A transition from a (111) oriented δ-TiN columnar grain structure to a nanocomposite type microstructure of equiaxed (200) oriented stoichiometric δ-TiN nanocrystalline grains embedded in a Ni rich amorphous phase was observed as Ni was incorporated into the film. This transition was accompanied by a complete relief of the compressive residual stressed induced by ion bombardment, presumably because the Ni rich amorphous phase that surrounds the TiN nanograins contributes to relieve part of these residual stresses.
- The elastic modulus of the Ti-N-Ni coatings remains constant for Ni contents up to 13 at. %, implying that the coatings are primarily formed by stoichiometric TiN and that the segregation of Ni does not affect its bonding strongly. Moreover, the Ni/TiN interfaces that develop are very strong and contribute to an important increase in hardness and fracture toughness with Ni content.
- A columnar grain structure with a more random texture develops for Ni contents higher than 13 at.% and the mechanical properties are drastically reduced. The reduction in mechanical properties is presumably due to a very defective TiN phase that develops due to the excess amount of Ni, that results in a highly over stoichiometric TiN columnar grains with voided columnar boundaries and the formation of Ni rich amorphous islands.

**Acknowledgements**

This investigation was supported by the European Research Council (ERC) under the European Union's Horizon 2020 research and innovation program (Advanced Grant VIRMETAL, grant agreement No. 669141). JMMA acknowledges funding from Comunidad de Madrid (IND2018/IND-9668). The authors wish to acknowledge J. Ferrer from the Centro Nacional de Aceleradores (CNA/CSIC) of Seville where the RBS measurements were carried out, as well as Miguel Monclús and Manuel Avella whose technical assistance and expertise have helped to make this article possible.**Data availability**

The datasets generated and/or analysed during the current study are available from the corresponding author on reasonable request.